\documentclass[letterpaper,10pt]{article} 

\usepackage{opticameet3} 

\newcommand\authormark[1]{\textsuperscript{#1}}

\usepackage{amsmath,amssymb}
\usepackage[colorlinks=true,bookmarks=false,citecolor=blue,urlcolor=blue]{hyperref} 

\begin{document}

\title{32$\mathrm{\times}$100 GHz WDM filter based on \\ultra-compact silicon rings with \\a high thermal tuning efficiency of 5.85 mW/$\mathrm{\pi}$}


\author{Qingzhong Deng,\authormark{1, *} 
Ahmed H. El-Saeed,\authormark{1, 2}
Alaa Elshazly,\authormark{1, 2}
Guy Lepage,\authormark{1}\\
Chiara Marchese,\authormark{1}
Hakim Kobbi,\authormark{1}
Rafal Magdziak,\authormark{1}
Jeroen De Coster,\authormark{1}\\
Neha Singh,\authormark{1}
Marko Ersek Filipcic,\authormark{1}
Kristof Croes,\authormark{1}
Dimitrios Velenis,\authormark{1}\\
Maumita Chakrabarti,\authormark{1}
Peter De Heyn,\authormark{1}
Peter Verheyen,\authormark{1}\\
Philippe Absil,\authormark{1}
Filippo Ferraro,\authormark{1}
Yoojin Ban,\authormark{1}
and Joris Van Campenhout\authormark{1}}

\address{\authormark{1} imec, Kapeldreef 75, 3001 Leuven, Belgium\\
\authormark{2} Photonics Research Group, Department of Information Technology, Ghent University-imec, Ghent, Belgium}

\email{\authormark{*}qingzhong.deng@imec.be} 

\copyrightyear{2024}

\begin{abstract}
To the best of our knowledge, this paper has achieved the lowest thermal tuning power (5.85 mW/$\mathrm{\pi}$) for silicon rings with FSR$\geq$3.2 THz, and the first silicon ring-based WDM-32$\mathrm{\times}$100 GHz filter.
\end{abstract}

\section{Introduction}

As the key to scale up optical communication capacity, wavelength-division multiplexing (WDM) has been extensively investigated in silicon photonics.
Silicon ring resonators, owing to their compact size and promising performance, have emerged as a focal point in WDM applications.
Despite their potential, existing silicon ring-based WDM filters are constrained, supporting a maximum of 16 channels~\cite{SiliconBasedChip_P2023a, SiMicroRing_OE2011b, Novel16Channel_2022a}.
\begin{figure}[!hb]
  \centering
  \includegraphics{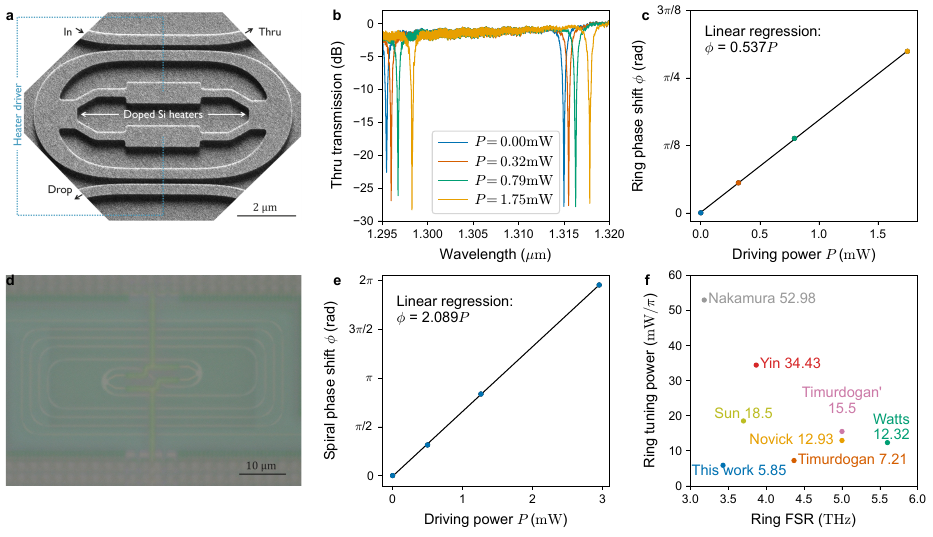}
  \caption{
  The scanning electron microscope (SEM) image (a), the measured Thru spectra under different driving powers (b), and the extracted thermal tuning power (c) of the ring resonator with doped silicon heaters.
  The SEM is taken after silicon etching process so that the electrical connections are not displayed.
  The optical microscope image (d), and the thermal tuning power (e) of the spiral phase shifter with doped silicon heaters.
  The ring tuning power comparison with literature (f): Nakamura~\cite{HighSpeedChip_AN2022a}, Yin~\cite{SiliconBasedChip_P2023a}, Sun~\cite{IntegratedMicroringTuning_2013a}, Timurdogan'~\cite{AdiabaticResonantMicroring_2013a}, Novick~\cite{LowLosswideFsr_2023a}, Watts~\cite{AdiabaticResonantMicrorings_2COLAEAQEALSCC2V12009a}, and Timurdogan~\cite{LShapedResonant_2013a}.
}
  \label{fig_heater_tuning}
\end{figure}
\begin{figure}[!hb]
  \centering
  \includegraphics{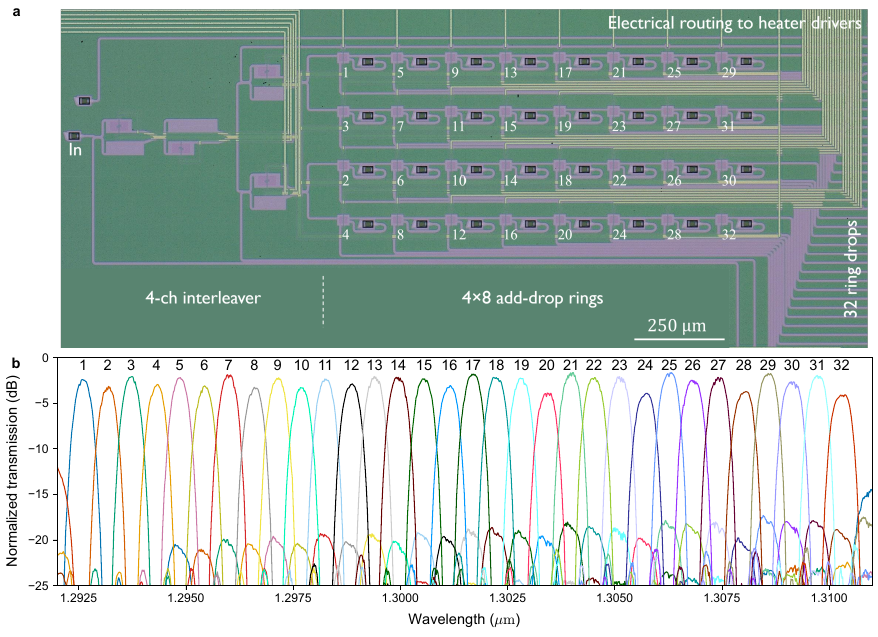}
  \caption{The optical microscope image (a) and the measured transmission spectra (b) of the proposed silicon ring-based WDM-32$\mathrm{\times}$100 GHz filter.}
  \label{fig_wdm32x100_spectra}
\end{figure}
To scale up to more WDM channels, a ring resonator with an ultra-compact form factor and remarkably low internal loss is imperative.
Recently, we have reported a low-loss waveguide bend supporting a whispering gallery mode which is experimentally proven to substantially reduce losses in circular bends from 0.372 dB to a mere 0.016 dB~\cite{LowLossWaveguide_2023a}.
Employing this bend design, a ring resonator with a 2 µm radius exhibits a free spectral range (FSR) of 3.77 THz, maintaining a low ring internal loss of 0.038 dB~\cite{LowLossWaveguide_2023a}.
In this paper, doped silicon heaters are implemented in the ring design, resulting in an exceptional thermal tuning efficiency 5.85 mW/$\mathrm{\pi}$.
Moreover, a WDM-32$\mathrm{\times}$100 GHz filter is demonstrated utilizing this ring design.

\section{Ring resonator with doped silicon heaters}
In the context of waveguide bend design, the strategic implementation of a smoothly varying width enables the excitation of a whispering gallery mode, confining light solely by the outer boundary~\cite{LowLossWaveguide_2023a}.
The inner boundary of these bends serves as an ideal space for integrating doped silicon heaters and connections to the heater driver, as visually depicted in Fig.~\ref{fig_heater_tuning}a.
This meticulous arrangement ensures that the doped silicon and metal wires remain non-overlapping with the light path within the ring, thereby preventing any additional ring loss.
Meanwhile, the heating process occurs within the silicon waveguide, enhancing thermal tuning efficiency.
The resulting ring spectra, obtained under varying driving power levels, are illustrated in Fig.~\ref{fig_heater_tuning}b. By tracking the resonance wavelength drift in these spectra, the corresponding ring phase shifts are extracted, as graphically represented in Fig.~\ref{fig_heater_tuning}c.
This approach yields a remarkable tuning power of 5.85 mW/$\mathrm{\pi}$ (calculated as $\pi/0.537$), surpassing existing literature results for rings with FSR$\geq$3.2 THz.
The FSR$\geq$3.2 THz is emphasized in the comparison because this is required for WDM-32$\mathrm{\times}$100 GHz, and the ring form factor is bounded by the FSR.
Notably, when the ultra-compact form factor is not a must, as in the Mach-Zehnder interferometer (MZI), the tuning power can be further reduced.
For instance, integrating doped silicon heater within a spiral waveguide, as demonstrated in Fig.~\ref{fig_heater_tuning}d, results in a tuning power of 1.50 mW/$\mathrm{\pi}$ (Fig.~\ref{fig_heater_tuning}e).

\section{Silicon ring-based WDM-32$\mathrm{\times}$100 GHz filter}
%
%
%

As shown in Fig.~\ref{fig_wdm32x100_spectra}a, a 32$\mathrm{\times}$100 GHz WDM filter is implemented with the proposed add-drop ring resonators and the spiral phase shifters.
The spiral phase shifters are used to construct a 4-channel MZI interleaver that divides the input 32$\mathrm{\times}$100 GHz signals into 4 groups of 8$\mathrm{\times}$400 GHz signals.
Each group is de-multiplexed by 8 add-drop rings coupled to the interleaver output waveguide.
All the doped silicon heaters in the interleaver and the rings are driven in parallel to align the working wavelengths.
As depicted by the measured spectra in Fig.~\ref{fig_wdm32x100_spectra}b, this WDM-32$\mathrm{\times}$100 GHz filter shows 3 dB bandwidth of 75$\pm$4 GHz, insertion loss of 1.64 $\sim$ 4.06 dB, and crosstalk of -29.1 $\sim$ -16.2 dB.
The filter tuning efficiency is estimated to be 283 GHz/mW per channel based on the individual characterization of the ring and the spiral phase shifter.
Compared to the results existing in literature (Tab.~\ref{tab_wdm_comparison}), this filter has competitive performance, and more importantly is the first silicon ring-based 32-channel WDM filter.
\begin{table}[!ht]
  \centering
  \caption{Performance comparison of silicon ring based WDM filters.}
  \begin{tabular}{cccccc}
    \hline
      Reference & Channel amount & Channel spacing & 3 dB bandwidth & Insertion loss & Cross-talk \\
      \ & \ & (GHz) & (GHz) & (dB) & (dB) \\ \hline
      \cite{SiliconBasedChip_P2023a} & 16 & 200 & - & 3.9 $\sim$ 5.6 &  -33.2 $\sim$ -23.2 \\ 
      \cite{SiMicroRing_OE2011b} & 16 & 100 & 70 & $\sim$5 & $\sim$ -20 \\ 
      \cite{Novel16Channel_2022a} & 16 & 80 & 32 & - & - \\  
      This work & 32 & 100 & 75$\pm$4 & 1.64 $\sim$ 4.06 & -29.1 $\sim$ -16.2 \\\hline
  \end{tabular}
  \label{tab_wdm_comparison}
\end{table}

\section{Conclusion}
In summary, this study presents a significant advancement in silicon photonics by integrating doped silicon heaters into waveguide bends.
The resulting add-drop ring resonator exhibits an impressive thermal tuning power of 5.85 mW/$\mathrm{\pi}$ while maintaining an FSR of 3.43 THz.
The tuning power to further reduced to 1.50 mW/$\mathrm{\pi}$ in a spiral phase shifter.
Building upon these achievements, we have successfully demonstrated a WDM filter encompassing 32 channels at 100 GHz spacing, showcasing competitive performance.
To the best of our knowledge, this study represents a pioneering effort, achieving the lowest thermal tuning power reported for silicon rings with FSR$\geq$3.2 THz.
Furthermore, this work marks the inception of the first 32-channel silicon ring-based WDM filter, underscoring the potential of our innovative approach in advancing the field of silicon photonics.
These findings will expand the horizons of silicon ring resonator, and also pave the way for enhanced and efficient photonic devices in high-capacity communication systems.

This work was supported by imec's industry-affiliation R\&D program “Optical I/O”.
\bibliographystyle{opticajnl}

\end{document}